\documentclass[twocolumn,showpacs,preprintnumbers,amsmath,amssymb]{revtex4-1}
\usepackage{amssymb}		
\usepackage{amsmath}
\usepackage{graphicx}
\usepackage[normalem]{ulem}
\usepackage{multirow}
\usepackage{appendix}
\usepackage{CJK}
\usepackage[usenames]{color}
\usepackage{bm}
\usepackage[colorlinks,linkcolor=blue,urlcolor=blue,citecolor=blue]{hyperref}
\usepackage{booktabs} 	
\usepackage{leftidx}

\makeatletter

\newcommand{\Rmnum}[1]{\expandafter\@slowromancap\romannumeral #1@}
\makeatother
\begin{document}
\begin{CJK*} {UTF8} {gbsn}

\title{System size dependence of baryon-strangeness correlations in relativistic heavy ion collisions from a multiphase transport model}

\author{Dong-Fang Wang(王东方)}

\author{Song Zhang(张松)}\thanks{Email: song\_zhang@fudan.edu.cn}

\author{Yu-Gang Ma(马余刚)}\thanks{Email:  mayugang@fudan.edu.cn}
\affiliation{Key Laboratory of Nuclear Physics and Ion-beam Application (MOE), Institute of Modern Physics, Fudan University, Shanghai 200433, China}

\begin{abstract}
The system size dependence of baryon-strangeness (BS) correlations ($C_{BS}$) are investigated with a multiphase transport (AMPT) model for various collision systems from 
$\mathrm{^{10}B+^{10}B}$, $\mathrm{^{12}C+^{12}C}$, $\mathrm{^{16}O+^{16}O}$, $\mathrm{^{20}Ne+^{20}Ne}$, $\mathrm{^{40}Ca+^{40}Ca}$, $\mathrm{^{96}Zr+^{96}Zr}$, and $\mathrm{^{197}Au+^{197}Au}$ at RHIC energies $\sqrt{s_{NN}}$ of 200, 39, 27, 20, and 7.7 GeV.
Both effects of hadron rescattering and a combination of different hadrons play a leading role for baryon-strangeness correlations. When the kinetic window is limited to absolute rapidity $|y|>3$, these correlations tend to be constant after the final-state interaction 
whatever kind of hadrons subset we chose based on the AMPT framework. The correlation is found to smoothly increase with baryon chemical potential $\mu_B$,  corresponding to the collision system or energy from the quark-gluon-plasma-like phase to the hadron-gas-like phase.
Besides, the influence of initial nuclear geometrical structures of $\alpha$-clustered nuclear collision systems of $\mathrm{^{12}C+^{12}C}$ as well as $\mathrm{^{16}O+^{16}O}$ collisions is discussed but the effect is found negligible.
The current model studies provide baselines for searching for the signals of Quantum Chromodynamics (QCD) phase transition and critical point in heavy-ion collisions through the BS correlation.
 	\end{abstract}
\maketitle

	\section{Introduction}
	\par
	
	Relativistic heavy-ion collisions create nuclear matter with sufficient energy density that one expects a quark-gluon plasma to form~\cite{phase_e1,phase_e2,phase_e3,phase_e4}.
	The fundamental challenge remains how to identify this hot and dense quark matter and fully understand the phase diagram of QGP matter.
	Lattice QCD calculations have indicated that the transition from  hadronic phase to  QGP phase is a crossover at zero baryon-chemical potential $(\mu_{B}=0)$ with a transition temperature $T_{c}\approx166$ MeV~\cite{tc_1,tc_2}.  
	For the finite size system, the transition temperature $T_{c}$
	could shift to a value higher than that in an unconstrained space \cite{Liu2020}.
In an attempt to address these considerations, researchers performed, from 2010 to 2017, a beam-energy scan  ~\cite{bes_1,bes_2,bes_3} 
	at  the BNL Relativistic Heavy Ion Collider (RHIC). One of the promising approaches to probe the QGP phase transition involves fluctuations~\cite{koch_re,jeon2003eventbyevent}.
	\par
	Theoretical calculations predicted that fluctuations and correlations of conserved charges were distinctly different in the hadronic or QGP phase~\cite{LuoNST}, 
	and they were experimentally accessible to distinguishing between these two phases~\cite{Adare_2016}.
	Experimental analysis of event-by-event fluctuations of net-conserved charges like baryon number ($B$), electric charge ($Q$) and strangeness ($S$), in particular, their higher-order cumulants were reported at RHIC~\cite{exp_1,exp_2} and LHC~\cite{exp_3,exp_4}.
	One of the event-by-event fluctuations observable was proposed by Koch~\cite{Koch_origin}, namely the baryon-strangeness correlation coefficient, 
	\begin{eqnarray}
	\begin{aligned}
	 C_{BS} = -3\frac{\left\langle BS \right\rangle - \left\langle B \right\rangle \left\langle S \right\rangle}{\left\langle S^{2}  \right\rangle - \left\langle S \right\rangle^{2}},
	 \end{aligned}
	\end{eqnarray}
	where $B$ and $S$ are the net-baryon number and net-strangeness in one event, respectively.
	The average value of $B$ and $S$ over a suitable ensemble of events is denoted by $\left\langle \cdot \right\rangle$.
	The $BS$ correlation was considered as a useful tool to characterize that the highly compressed and heated matter created in heavy-ion collisions underwent ideal QGP phase,  strongly coupled QGP phase, or hadronic phase.
In previous analyses,  several specific models were applied, such as $(2+1)$ Polyakov Quark Meson Model~\cite{u_model1}, 
	hadron resonance gas model~\cite{u_model2,u_model3}, UrQMD~\cite{u_model4,Haussler_2007,Luo_UrQMD} model as well as the AMPT model~\cite{Jin_2008}
	 to study the fluctuations and compare them with Lattice QCD results~\cite{lattice_1,lattice_2}.
	\par

Research on small systems has been performed for several  years, both experimentally and theoretically~\cite{small_sys_LHC,small_sys_1}, and several proposals for a  system scan (e.g. $\mathrm{O}+\mathrm{O}$)~\cite{PRC2029sysScanLHC} have been made to study the possible signals of QGP matter in small systems as well as to investigate the initial state effects on the final state observables~\cite{SHuang2020sysScan,PRC2029sysScanLHC,ZHANG2020135366}.
	We noticed that the ALICE collaboration reported the enhanced production of multi-strange hadrons in high-multiplicity proton-proton collisions~\cite{smallSystemALICE2017}. In the same context, we consider the baryon-strangeness correlation which is related to the QGP phase transition may also be sensitive to the fluctuations from small systems to large systems through heavy-ion collisions.

	\begin{table*}[]
\scriptsize
\centering
\caption{AMPT input parameters and $\left\langle \mathrm{N_{part}}\right\rangle$ values of different collision systems.}
\label{AMPT_info}

	\begin{tabular}{cc|cc|cc|cc}
\toprule
\multicolumn{2}{c}{} & \multicolumn{2}{c}{$\sqrt{s_{NN}}$ = 200 GeV} & \multicolumn{2}{c}{$\sqrt{s_{NN}}$ = 20 GeV} & \multicolumn{2}{c}{$\sqrt{s_{NN}}$ = 7.7 GeV} \\
\cmidrule(r){3-4} \cmidrule(r){5-6} \cmidrule(r){7-8}
System & $\mathrm{\it{b_{\text{max}}}}$[fm]
&$\left\langle \mathrm{N_{part}}\right\rangle$ &Event counts
&$\left\langle \mathrm{N_{part}}\right\rangle$ &Event counts
&$\left\langle \mathrm{N_{part}}\right\rangle$ &Event counts      \\
\hline
$\mathrm{\leftidx{^{10}}B} + \mathrm{\leftidx{^{10}}B}$		&1.15619		&14.8  &7$\times 10^{4}$    	&13.2  &12$\times 10^{4}$  &13.1  &16$\times 10^{4}$\\
$\mathrm{\leftidx{^{12}}C} + \mathrm{\leftidx{^{12}}C}$		&1.22864		&18.7  &10$\times 10^{4}$    	&16.8  &6$\times 10^{4}$ 	&16.7  &10$\times 10^{4}$\\
$\mathrm{\leftidx{^{16}}O}+\mathrm{\leftidx{^{16}}O}$		&1.35229		&25.5   &10$\times 10^{4}$	&23.1  &4$\times 10^{4}$	&23.0  &10$\times 10^{4}$ \\
$\mathrm{\leftidx{^{20}}Ne}+\mathrm{\leftidx{^{20}}Ne}$		&1.45671		&32.8  &2$\times 10^{4}$	&30.0  &4$\times 10^{4}$	&29.8  &2$\times 10^{4}$\\
$\mathrm{\leftidx{^{40}}Ca}+\mathrm{\leftidx{^{40}}Ca}$		&1.83534		&69.3  &2$\times 10^{4}$	&65.0  &1$\times 10^{4}$	&64.9  &1$\times 10^{4}$ \\
$\mathrm{\leftidx{^{96}}Zr}+\mathrm{\leftidx{^{96}}Zr}$ 		&2.45727		&174.2 &2$\times 10^{4}$	&167.3  &2$\times 10^{4}$	&166.9 &3$\times 10^{4}$\\
$\mathrm{\leftidx{^{197}}Au}+\mathrm{\leftidx{^{197}}Au}$		&3.1226		&364.1 &1$\times 10^{4}$	&354  &3$\times 10^{4}$	&353.8  &3$\times 10^{4}$\\
\bottomrule
\end{tabular}
\end{table*}

	\par
	In this work, we adopt a multiphase transport model to study the influence of collision system size on baryon-strangeness correlations $C_{BS}$.
By tuning the collision energies of two nuclei, we investigate the energy dependence of correlations $C_{BS}$.
	The maximum rapidity acceptance $y_{\text{max}}$ dependence and the influence of initial nuclear geometry structure are also discussed.
	
	The paper is organized as follows. 
	First, in Sec.\ref{sec:model},  a short introduction to a multiphase transport model and some input parameters are presented
	and the physical picture of baryon-strangeness correlations is briefly manifested.
	Next, based on AMPT model, the dependence of baryon-strangeness correlations as a function of system size, center-of-mass energy, and $y_{\text{max}}$ are discussed in Sec.\ref{sec:result}. 
	Finally, a brief summary is presented in Sec.\ref{sec:summary}.

	\section{Model and methodology}
	\label{sec:model}
	\subsection{Brief introduction to AMPT model}
	
	A multi-phase transport model (AMPT), which is a hybrid dynamic model, is employed to calculate different collision systems.
	The AMPT model can describe the $p_{T}$ distribution of charged particles \cite{xujun,suppressionhighpt,Ye_2017,JinXH} and their elliptic flow of Pb+Pb collisions
	at $\sqrt{s_{NN}} = 2.76$ TeV, as measured through the LHC-ALICE Collaboration. The model includes four main components to describe
	the relativistic heavy ion collision process: the initial condition which is simulated using the Heavy Ion Jet Interaction Generator (HIJING) model~\cite{HIJING-1,HIJING-2}, the partonic interaction which is described by Zhang's Parton Cascade (ZPC) model~\cite{ZPCModel}, the hadronization process which goes through by a Lund string fragmentation or coalescence model, and the hadron rescattering process which is treated by A Relativistic Transport (ART) model~\cite{ARTModel}. There are two versions of AMPT: 1) the AMPT version with a string melting mechanism, in which a partonic phase is generated from excited strings in the HIJING model, where a simple quark coalescence model is used to combine the partons into hadrons; and 2) the default AMPT version which only undergoes a pure hadron gas phase. 
	The AMPT model succeeds to describe extensive physics topics for relativistic heavy-ion collisions at the RHIC~\cite{AMPT_origin} as well as the LHC~\cite{AMPTGLM2016} energies, 
e.g. for hadron HBT correlation~\cite{AMPTHBT}, di-hadron azimuthal correlation~\cite{AMPTDiH,WangHai}, collective flows~\cite{STARFlowAMPT,AMPTFlowLHC}, strangeness productions~\cite{JinXH,SciChinaJinS} as well as chiral magnetic effects and so on \cite{Zhao,Huang,Wang,XuZW}.
The details of AMPT can be found in Ref.~\cite{AMPT_origin}.
		In the AMPT model, impact parameter $b$, demonstrating the transverse distance between the centers of the two collided nuclei, can determine the collision centrality. The number of participants $ \mathrm{N_{part}}$ is also related to the centrality or impact parameter. We adopt the AMPT parameters as suggested in Ref.~\cite{Ye_2017}. The calculated collision systems and energies, their corresponding maximum impact parameters, $ \mathrm{N_{part}}$ and the number of events are listed in Table~\ref{AMPT_info}.
		
	\subsection{Baryon-strangeness correlations}
	
		Finding a suitable probe to distinguish QGP matter is the key to understanding the QGP phase transition in relativistic heavy-ion collisions.
		The correlation coefficient $C_{BS}$, calculated via conserved quantities which are less affected due to uncertainty from hadronization, has an advantage over other probes. 
	 Under ideal QGP assumption, where the basic degrees of freedom are weakly interacting quarks and gluons at high temperature,  $\left\langle S \right\rangle$ keeps $0$ and $C_{BS}$ can be written as $C_{BS} = -3 \frac{\left\langle BS \right\rangle}{\left\langle S^{2}\right\rangle}=1$,
	noting that the strangeness is only carried by $s$ quark~\cite{Koch_origin}.
	However, this feature is different from a hadron gas phase where this coefficient strongly depends on the hadronic environment.
	 Based on an assumption of uncorrelated multiplicities, $C_{BS}$ can be written as~\cite{Koch_origin}
\begin{equation}
C_{B S} \approx 3 \frac{\langle\Lambda\rangle+\langle\bar{\Lambda}\rangle+\cdots+3\left\langle\Omega^{-}\right\rangle+3\left\langle\bar{\Omega}^{+}\right\rangle}{\left\langle K^{0}\right\rangle+\left\langle\bar{K}^{0}\right\rangle+\cdots+9\left\langle\Omega^{-}\right\rangle+9\left\langle\bar{\Omega}^{+}\right\rangle}.
\label{eq_CBS_Approx}
\end{equation}
	In actual calculation~\cite{Koch_origin}, $C_{BS}$ is expressed as
\begin{equation}
C_{B S} = -3 \frac{\sum_{n} B^{(n)} S^{(n)}-\frac{1}{N}\left(\sum_{n} B^{(n)}\right)\left(\sum_{n} S^{(n)}\right)}{\sum_{n}\left(S^{(n)}\right)^{2}-\frac{1}{N}\left(\sum_{n} S^{(n)}\right)^{2}},
\end{equation}
where $B$ and $S$ denote the net baryon number and net strangeness observed for a given event, respectively, and 
$N$ is the total number of events. Furthermore, some attention should also be paid to 
the statistical errors as suggested in Refs.~\cite{Luo_UrQMD,Luo_2012} (see the Appendix).

	\section{Results and discussion}
	\label{sec:result}
\begin{figure}[htb]
				\includegraphics[angle=0,scale=0.45]{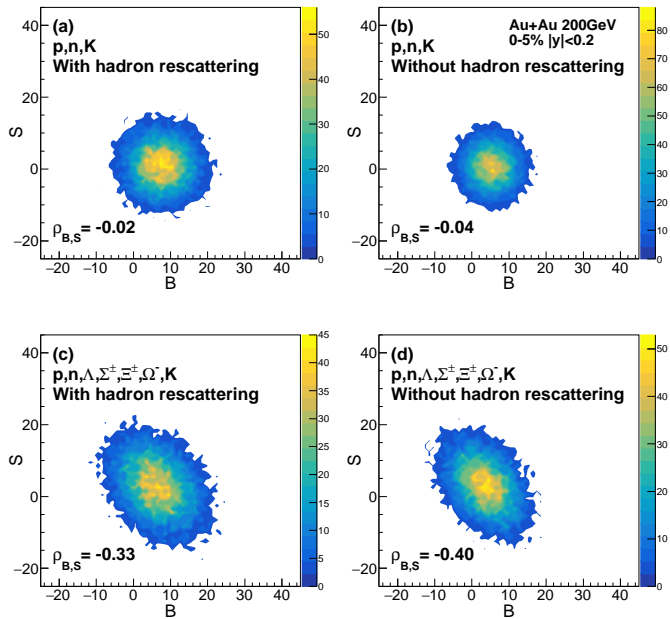}
				\caption{The correlation between net-baryon $B$ and net-strangeness $S$ for two different subsets of hadrons in the most central (0$-$5\%) $\mathrm{^{197}Au+^{197}Au}$ collisions at $\sqrt{s_{NN}} = 200$ GeV with the string melting AMPT framework. Panel (a) and (b) correspond to the Case (II) w/ and w/o hadron rescattering, while  (c) and (d) correspond to the Case (I) w/ and w/o hadron rescattering.}
				\label{Fig1_BS_plane}
	\end{figure}

		The combination of hadrons would play an important role in the measurement of the baryon-strangeness correlation $C_{BS}$. To investigate this effect, the distribution of net-baryon $B$ versus net-strangeness $S$ was presented in Fig.~\ref{Fig1_BS_plane}.
		We chose two combinations of hadrons for baryon-strangeness correlations calculation in this figure:
		Case (\Rmnum{1}) $p$+$n$+$\Lambda$+$\Sigma^{\pm}$+$\Xi^{\pm}$+$\Omega^{-}$+$K$ and Case (\Rmnum{2}) $p$+$n$+$K$, 
	where both the particles and anti-particles were included with kinetic windows $0.1<p_{T}<3.0$ GeV/$\textit{c}$ and $|y|<0.2$.
	In this figure, we observed that the baryon-strangeness distribution was more concentrated on the center if  the effect of hadron rescattering is off, which contributes to stronger correlations.  Once counting more strange baryons (and anti-baryons), the distribution would be stretched into elliptical distribution and leads to finite negative correlations. This correlation  can be represented as the following:
	\begin{equation}
\rho_{B, S}=\frac{\operatorname{cov}(B, S)}{\sigma_{B} \sigma_{S}}=\frac{\langle(B-\langle B\rangle)(S-\langle S\rangle)\rangle}{\sqrt{\left\langle B^{2}\right\rangle-\langle B\rangle^{2}} \sqrt{\left\langle S^{2}\right\rangle-\langle S\rangle^{2}}}.
\end{equation}
The more strange baryons were used, the more negative correlation was presented between $B$ and $S$.
	
	 We would focus on hadron combination case (\Rmnum{2}) for calculating correlations,  and also present case (\Rmnum{1}) results for comparing effects from different hadrons combinations.
	 
\begin{figure*}[htb]
				\includegraphics[angle=0,scale=0.85]{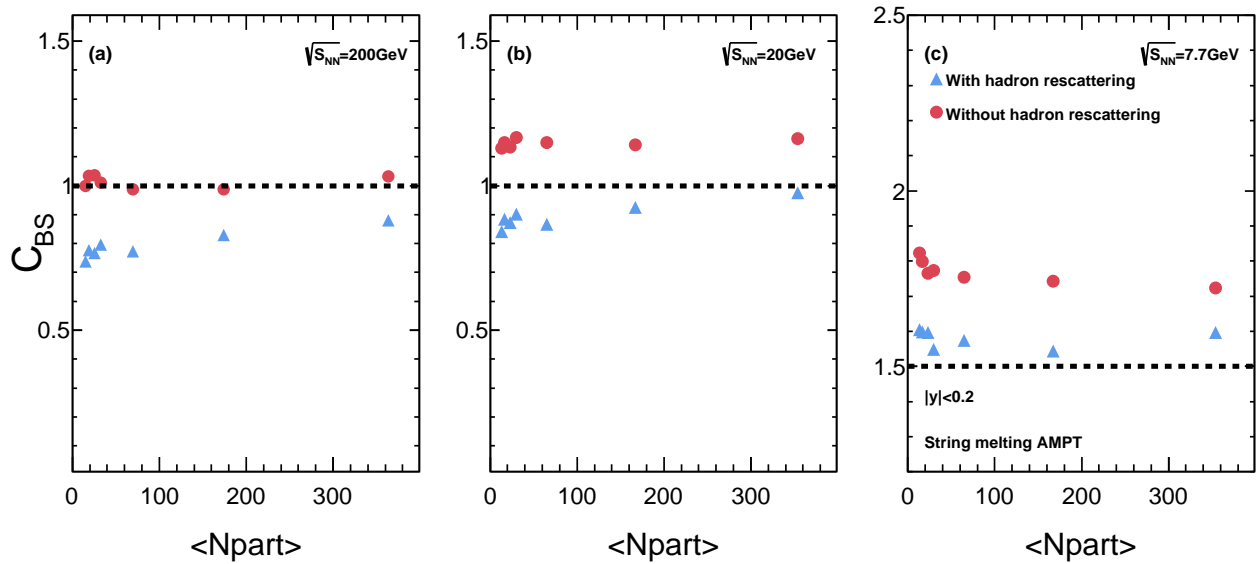}
				\caption{The baryon-strangeness correlation $C_{BS}$ versus $\left\langle \mathrm{N_{part}}\right\rangle$ at  $\sqrt{s_{NN}}$ = 200, 20 and 7.7 GeV in the most central collisions (0$-$5\%) of $\mathrm{^{10}B+^{10}B}$, $\mathrm{^{12}C+^{12}C}$, $\mathrm{^{16}O+^{16}O}$, $\mathrm{^{20}Ne+^{20}Ne}$, $\mathrm{^{40}Ca+^{40}Ca}$, $\mathrm{^{96}Zr+^{96}Zr}$, and $\mathrm{^{197}Au+^{197}Au}$ systems at RHIC energies $\sqrt{s_{NN}}$ = 200 (a), 20 (b), and 7.7 GeV (c) in the string melting AMPT framework.				}
				\label{Fig2_Sys_E_scan_CBS}
	\end{figure*}

	  Figure~\ref{Fig2_Sys_E_scan_CBS} shows the system size dependence for all particles (\Rmnum{2}) under the effect with/without hadron rescattering in 0-5\% $\mathrm{^{10}B+^{10}B}$, $\mathrm{^{12}C+^{12}C}$, $\mathrm{^{16}O+^{16}O}$, $\mathrm{^{20}Ne+^{20}Ne}$, $\mathrm{^{40}Ca+^{40}Ca}$, $\mathrm{^{96}Zr+^{96}Zr}$, and $\mathrm{^{197}Au+^{197}Au}$ collisions at $\sqrt{s_{NN}}$= 200  (a), 20  (b), and 7.7 (c) GeV from the AMPT model. 
	In the case without hadron rescattering where the hadronized system just experienced a partonic phase, the baryon-strangeness correlation $C_{BS}$ keeps constant at 200 and 20 GeV as collision system size increases. At $\sqrt{s_{NN}}$ = 7.7 GeV, $C_{BS}$ almost keeps a constant but displays a slightly decreasing trend with system size. As collision energy increases, $C_{BS}$ approaches the value conformed to an ideal QGP assumption ($C_{BS}$ = 1).  If hadron rescattering was taken into account  the baryon-strangeness correlation $C_{BS}$ exhibits similar behavior at 200 and 20 GeV, while it is not completely flat at 7.7 GeV. The rescattering process would erase the signal of partonic matter which is consistent with the earlier AMPT study~\cite{Jin_2008}. 
	
\begin{figure*}[htb]
				\includegraphics[angle=0,scale=0.90]{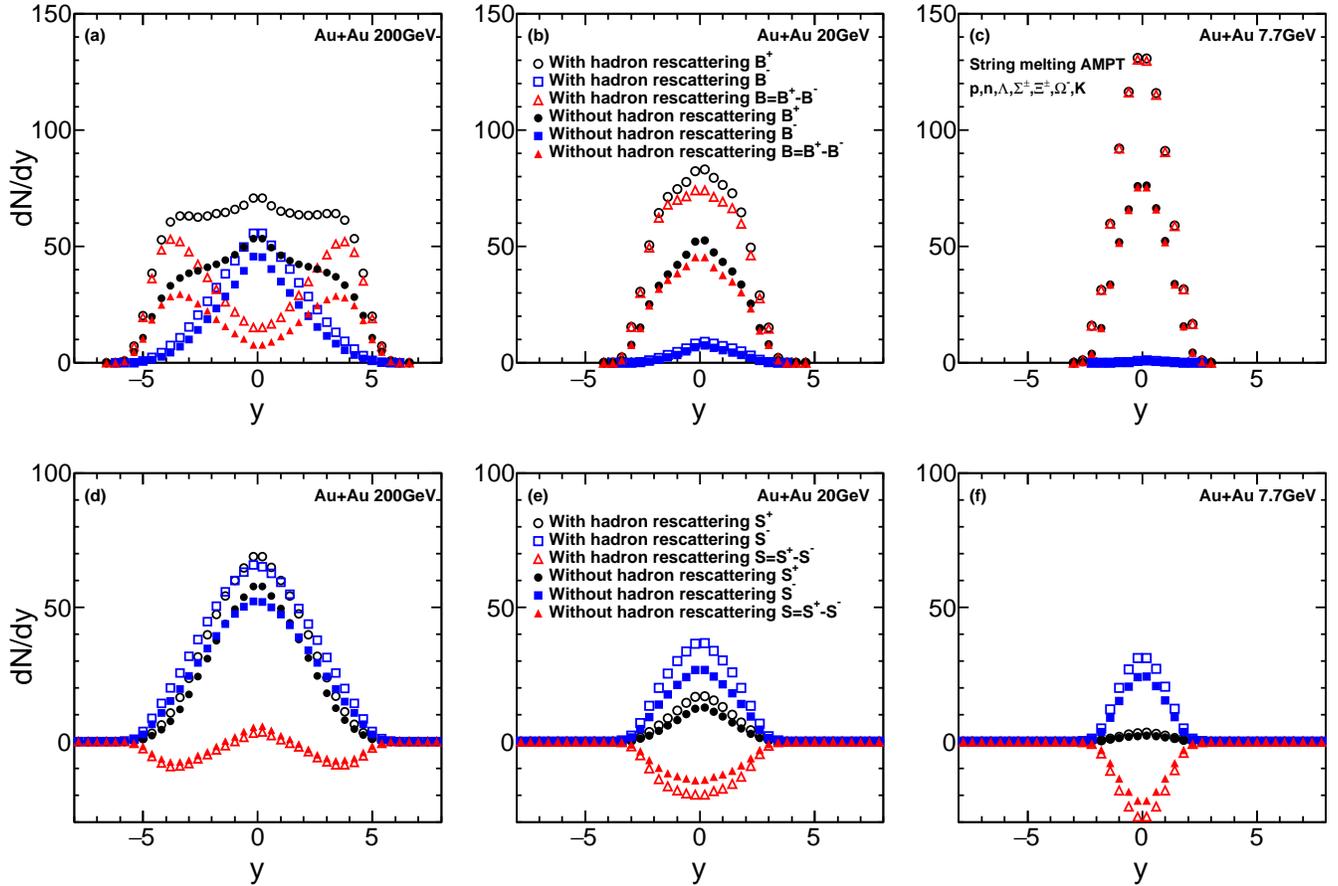}
				\caption{The 
				AMPT results of positive baryon (strangeness) $B^{+}$ ($S^{+}$), negative baryon (strangeness) $B^{-}$ ($S^{-}$) and net-baryon (-strangeness) $B$ ($S$) $dN/dy$ for identified particles, namely $p$, $n$, $\Lambda$, $\Sigma^{\pm}$, $\Xi^{\pm}$, $\Omega^{-}$, and  $K$ in
				$\mathrm{^{197}Au+^{197}Au}$ collisions at RHIC energies $\sqrt{s_{NN}}$ = 200 (a,d), 20 (b,e), and 7.7 (c,f) GeV based on the string melting AMPT framework. The meaning of different symbols are illustrated in the insert of (b) and (e). Here the kinematics window is $|y|<0.2$.
				}
				\label{Fig3_dNdy}
	\end{figure*}

	This dependence is also related to rapidity distribution, thus the baryon and strangeness yield $dN/dy$ (rapidity density) were also presented for $\mathrm{^{197}Au+^{197}Au}$ collisions at RHIC energies $\sqrt{s_{NN}}$ = 200, 20, and 7.7 GeV based on the string melting AMPT model, as shown in Fig.~\ref{Fig3_dNdy}. We observed that the rapidity  distribution of net-baryon $B$ becomes more concentrated in the middle rapidity when collision energy decreases as presented in Fig.~\ref{Fig3_dNdy}(a), (b), and (c). At $\sqrt{s_{NN}}$ = 20 and 7.7 GeV, the lower collision energy makes positive baryon $B^{+}$ much larger than negative baryon $B^{-}$ (almost contributed by anti-proton).
After the hadron rescattering process, as a result of decay particles contribution, net-baryon $B$ shows a little higher than the values without hadron rescattering. The non-Gaussian distribution of the $B$ rapidity at 200 GeV has also been found in Ref.~\cite{Lin_2017xkd}.

 Figure~\ref{Fig3_dNdy}(e) and (f) display the rapidity distribution of net-strangeness $S$ at $\sqrt{s_{NN}}$ =  20 and 7.7 GeV, respectively, where $S$ is always negative.
 However, $S$ turns to positive at 200 GeV in mid-rapidity as shown in Fig.~\ref{Fig3_dNdy}(d).
As energy increases, the rapidity density of $B$ decreases in the chosen region, and the baryon-strangeness correlation $C_{BS}$ is closer to 1, manifesting the system is close to the QGP state. 
The more net-baryon $B$ the collision system has, the state would be closer to the hadron gas phase with larger value of $C_{BS}$.

As energy decreases, the rapidity densities of net-baryon $B$ and net-strangeness $S$ are growing a more sharp peak as plotted in Fig.~\ref{Fig3_dNdy}.
Therefore, the final result of $C_{BS}$ will be affected greatly by a slightly changing of the dynamic window size.
Consequently, we could draw a conclusion that $C_{BS}$ is also affected by kinetic windows due to this non-flat rapidity distribution.

\par

	\begin{figure}[htb]
				\includegraphics[angle=0,scale=0.43]{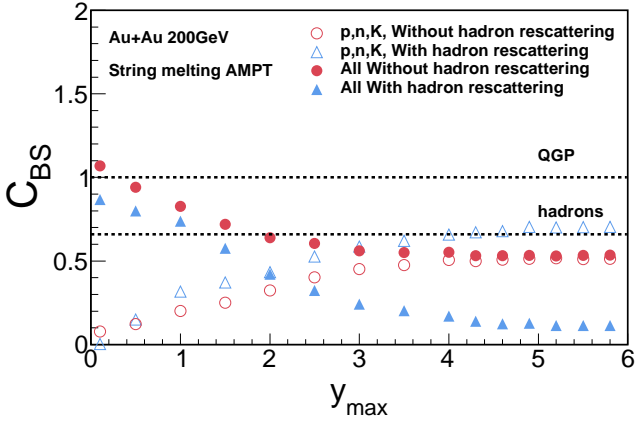}
				\caption{
				The maximum rapidity acceptance ($|y|<y_{\text{max}}$) dependence of the correlation coefficient $C_{BS}$
				in $\mathrm{^{197}Au+^{197}Au}$ collisions at $\sqrt{s_{NN}} = 200$ GeV in the string melting AMPT framework.
				Two different subsets of hadrons are adopted to show different dependencies.
				The two dashed lines indicate theoretical estimate of simple QGP ($C_{BS}$ = 1) and hadron gas ($C_{BS}$ = 0.66) at chemical freeze-out condition of $T$ = 170 MeV and $\mu_{b}$ = 0, respectively.  }
				\label{Fig4_ymax}
	\end{figure}
	
				\begin{figure}[htb]
				\includegraphics[angle=0,scale=0.43]{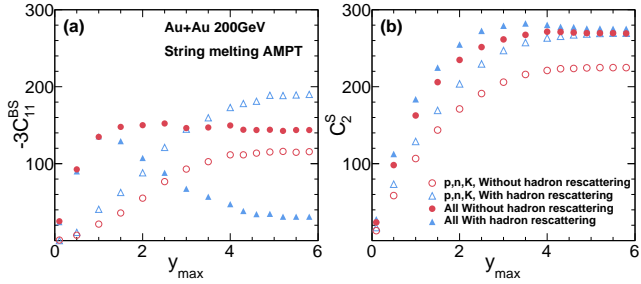}
				\caption{
				(a) The maximum rapidity acceptance ($|y|<y_{\text{max}}$) dependence of the numerator ($C_{11}^{BS}$) and (b) the denominator ($C_{2}^{S}$) (b) of $C_{BS}$
				in $\mathrm{^{197}Au+^{197}Au}$ collisions at $\sqrt{s_{NN}} = 200$ GeV with the string melting AMPT framework.
				Two different subsets of hadrons are adopted to show different dependencies.
				}
				\label{Fig5_Num_Den}
	\end{figure}
	
		\begin{figure}[htb]
				\includegraphics[angle=0,scale=0.44]{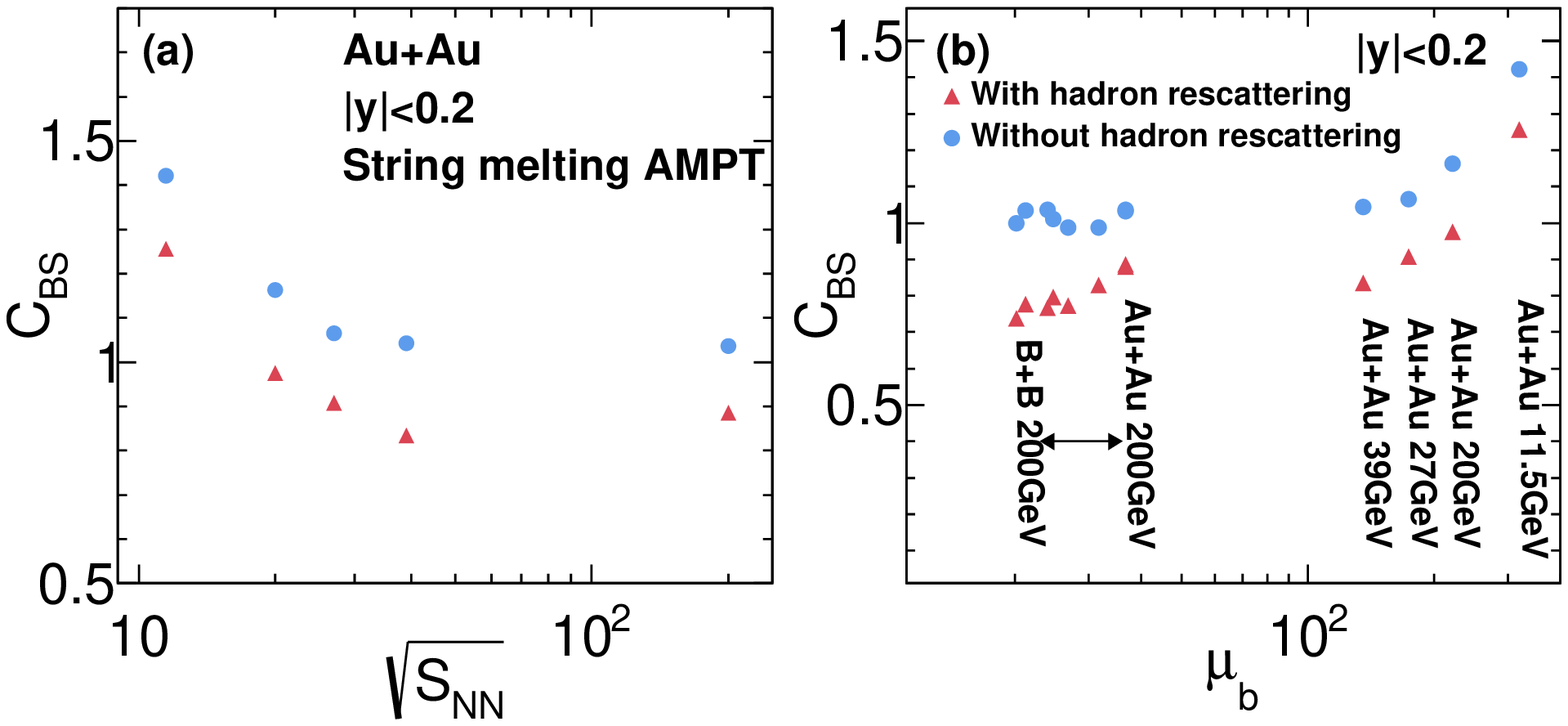}
				\caption{
				(a) The correlation coefficient $C_{BS}$ in the most central (0$-$5\%) $\mathrm{^{197}Au+^{197}Au}$ collisions is shown as a function of $\sqrt{s_{NN}}$; 
				(b) The correlation coefficient $C_{BS}$ for a hadron gas at freeze-out is shown as a function of the baryon chemical potential $\mu_{B}$ in the most central (0$-$5\%) collision at different collision systems and energy.
				For the $C_{BS}$ system scan, we chose
				 $\mathrm{^{10}B+^{10}B}$, $\mathrm{^{12}C+^{12}C}$, $\mathrm{^{16}O+^{16}O}$, $\mathrm{^{20}Ne+^{20}Ne}$, $\mathrm{^{40}Ca+^{40}Ca}$, $\mathrm{^{96}Zr+^{96}Zr}$, and $\mathrm{^{197}Au+^{197}Au}$ collisions at $\sqrt{s_{NN}}$ = 200 GeV.
				 For the $C_{BS}$ energy scan, the choice of energy is $\sqrt{s_{NN}}$ = 11.5, 20, 27, 39, and 200 GeV.
				}
				\label{Fig6_E_sys_mub}
	\end{figure}

	We also presented $C_{BS}$ as a function of the rapidity acceptance range $y_{\text{max}}$ in Au+Au collisions at $\sqrt{s_{NN}}$ = 200 GeV by the AMPT model.
	As manifested in Fig.~\ref{Fig4_ymax}, we observed two different $y_{\text{max}}$ dependence based on the two different combinations of hadrons.
	The correlation coefficient tends to increase with $y_{\text{max}}$ in the Case (\Rmnum{2}) hadrons combination.
	However, when the Case (\Rmnum{1}) hadrons combination was chosen, the coefficient tends to decrease.
	Although hadrons combination was different, $C_{BS}$ goes asymptotically to a constant as  $y_{\text{max}}>3$.
	Additionally, the choice of hadron combination has no effect on the result at large $y_{\text{max}}$ at the hadron rescattering stage as a consequence of conserved quantities of baryon number and strangeness. In previous study~\cite{Koch_origin}, the correlation coefficient $C_{BS}$ first increases with $y_{\text{max}}$  
	and reaches a maximum value at a certain $y_{\text{max}}$ before it drops to 0.

To understand this phenomenon, Fig.~\ref{Fig5_Num_Den}(a) and (b) display  the maximum rapidity acceptance ($|y|<y_{\text{max}}$) dependence of the numerator ($C_{11}^{BS}= \left\langle BS \right\rangle - \left\langle B \right\rangle \left\langle S \right\rangle$) and the denominator ($C_{2}^{S}= \left\langle  S^{2} \right\rangle -\left\langle S \right\rangle^{2}$) of $C_{BS}$ in $\mathrm{^{197}Au+^{197}Au}$ collisions at $\sqrt{s_{NN}} = 200$ GeV with the string melting AMPT model, respectively. In the case of (\Rmnum{2}) hadrons combination, both the $-3C_{11}^{BS}$ and the $C_{2}^{S}$ gradually tend to a constant value as $y_{\text{max}}$ increases as shown in Fig~\ref{Fig5_Num_Den}(a) and (b), respectively. However, in the case of (\Rmnum{1}) hadrons combination, the $-3C_{11}^{BS}$ increases with $y_{\text{max}}$ and then drops to a constant value with the hadron rescattering process. Thus, the value of $C_{11}^{BS}$ is the dominant factor affecting the correlation coefficient $C_{BS}$.

	\par
	The $C_{BS}$ calculated in this model was plotted in Fig.~\ref{Fig6_E_sys_mub}(a) at $\sqrt{s_{NN}}$ = 11.5, 20, 27, 39, and 200 GeV in $\mathrm{^{197}Au+^{197}Au}$ central collisions and presented strong energy dependence. As energy increases, $C_{BS}$ goes down to  0.8 at the top RHIC energy. This result was also below the expected value for an ideal QGP phase which was mentioned in Ref.~\cite{Haussler_2007}. Figure~\ref{Fig6_E_sys_mub}(b) shows $C_{BS}$ as a function of baryon chemical potential $\mu_{B}$ at chemical freeze-out for $\sqrt{s_{NN}}$ = 200 GeV collision systems, where   $\mu_{B}$ was extracted based on the thermal model as given in our previous paper~\cite{wdf}.
	At given collision energy, $\mu_{b}$ increases with system size, and a similar trend  also appears in Fig.~\ref{Fig2_Sys_E_scan_CBS}(a).
	The correlation coefficient $C_{BS}$ with the hadron rescattering process slightly increases with $\mu_{b}$, which is consistent with the previous conclusion~\cite{Koch_origin}.
	Meanwhile, the collision energy dependence of $C_{BS}$ was  displayed in Fig.~\ref{Fig6_E_sys_mub}(b).
	For a given system,  there were more net baryons in the collision system as energy decreases, leading to the $C_{BS}$ enhancement. The correlation coefficient presents a smooth baryon chemical potential dependence if collision systems and collision energies are characterized by the potential.
	\par
\begin{figure}[htb]
		\includegraphics[angle=0,scale=0.42]{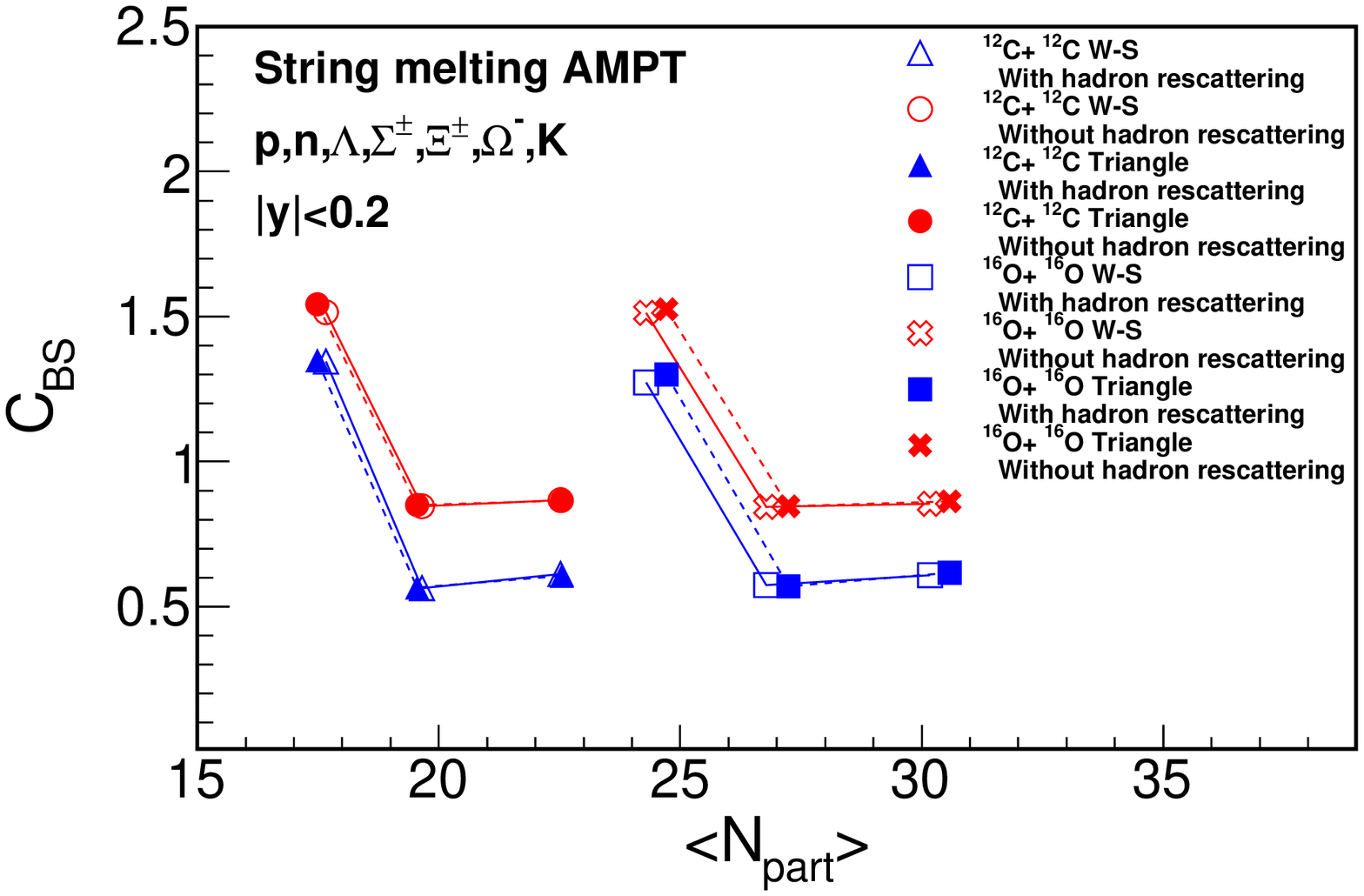}
					\caption{The correlation coefficient $C_{BS}$ as a function of the number of participants $\langle \mathrm{N_{part}}\rangle$ which are obtained from different  center-of-mass energy at $\sqrt{s_{NN}}$ = 10, 200, and 6370 GeV 
				in the most central collisions  (impact parameter $b=0$) of $\mathrm{^{12}C+^{12}C}$ and $\mathrm{^{16}O + ^{16}O}$ systems.}
				\label{Fig7}
	\end{figure}

	Finally, we investigated the possible effect of $\alpha$-clustering structure of light nuclei on the correlation coefficient $C_{BS}$.
	Some previous works proposed  the signatures of $\alpha$-clustering structure in light nuclei could be observed via heavy-ion collisions at ultra-relativistic energies~\cite{a_cluster_origin,AlphaModelHe1,AlphaClusterHIC-1,PhysRevC95064904,Zhang2,ChengYL,PhysRevC99044904,SHuang2020sysScan,HeJJ,MaLong}. In this context, we examined the influence of the fluctuation of  initial nuclear geometry structure on the correlation coefficient $C_{BS}$. To this end, the nucleon distribution might be considered as a three-$\alpha$ clustering triangle structure for $^{12}\mathrm{C}$ and  four-$\alpha$ clustering tetrahedron structure for $^{16}\mathrm{O}$ in the present study.  Fig.~\ref{Fig7} demonstrates  $\alpha$-clustering effects on $C_{BS}$ for $\mathrm{^{12}C+^{12}C}$ as well as $\mathrm{^{16}O + ^{16}O}$ collisions at $\sqrt{s_{NN}}$ = 10, 200, and 6370 GeV, respectively. These collision systems with different $\alpha$-clustering structure and energies were labeled by $\left\langle \mathrm{N_{part}}\right\rangle$. The results show  nearly no difference visibly between the Woods-Saxon nucleon distribution and the $\alpha$-clustering structures  via $C_{BS}$ coefficient.  It implies that the baryon-strangeness correlation was insensitive to initial nucleon distribution, which could in turn help to isolate other ingredients for affecting $C_{BS}$, such as the hadron rescattering effect as discussed in this work.  
	
\section{summary}
\label{sec:summary}	

In summary, we studied the system and energy dependence of the baryon-strangeness correlation coefficient in the framework of the AMPT model. 
The hadron rescattering process partly weaken the baryon-strangeness correlation as expected and weak decay contributions for strangeness or the count of baryons might have an effect on final results  which need to be further investigated.
The combination of different hadrons additionally affects the results significantly. Besides, it was found when the maximum rapidity acceptance $y_{\text{max}}>3$, the baryon-strangeness coefficient is independent of the combination of different hadrons in the final state based on the AMPT model. 
 The correlation coefficient could be grouped if the collision systems and collision energies were characterized by the baryon chemical potential.
 In addition, we investigated the effect of initial nucleon distribution for light nuclei, specifically with either the Woods-Saxon nucleon distribution or the  $\alpha$-clustering structure for  $^{12}$C and $^{16}$O nuclei, on the final baryon-strangeness correlation results but found negligible effect.  These AMPT model studies provide baselines for searching for the signals of QCD phase transition and critical point in heavy-ion collisions through the BS correlation.

\begin{acknowledgements}
	
This work was supported in part by Guangdong Major Project of Basic and Applied Basic Research No. 2020B0301030008, the Strategic Priority Research Program of CAS under Grant No. XDB34000000, the National Natural Science Foundation of China under contract Nos.  11875066, 11890710, 11890714, 11925502, 11961141003, National Key R\&D Program of China under Grant No. 2016YFE0100900 and 2018YFE0104600, and the Key Research Program of Frontier Sciences of the CAS under Grant No. QYZDJ-SSW-SLH002.

\end{acknowledgements}

	\section{Appendix}
		  \begin{appendices} 
	\section{observable} 
	The joint cumulant of several random variables $X_{1}, ..., X_{n}$ is defined by a similar cumulant generating function
	\begin{equation}
K\left(t_{1}, t_{2}, \ldots, t_{n}\right)=\log E\left(\mathrm{e}^{\sum_{j=1}^{n} t_{j} X_{j}}\right).
\end{equation}
	 A consequence is that
	 \begin{equation}
\kappa\left(X_{1}, \ldots, X_{n}\right)=\sum_{\pi}(|\pi|-1) !(-1)^{|\pi|-1} \prod_{B \in \pi} E\left(\prod_{i \in B} X_{i}\right)
\end{equation}
where $\pi$ runs through the list of all partitions of $\{ 1, ..., n \}$, $B$ runs through the list of all blocks of the partition $\pi$, and $|\pi|$ is the number of parts in the partition.
In this analysis, we use $B$ and $S$ to represent the net-baryon number and net-strangeness in one event, respectively.
The deviation of $B$ and $S$ from their mean value are expressed by $\delta B = B - \left\langle B \right\rangle$ and $\delta S = S -\left\langle S \right\rangle$, respectively. 
As mentioned above, we use $\left\langle \cdot \right\rangle$ to represent expected value.
 According Eq.(\ref{eq_CBS_Approx})
\begin{equation}
\begin{aligned}
C(\delta B,\delta S) &=  \left\langle \delta B \delta S \right\rangle = \left\langle BS \right\rangle - \left\langle B \right\rangle \left\langle S \right\rangle\\
C(\delta S,\delta S) &=  \left\langle \delta S \delta S \right\rangle = \left\langle S^{2} \right\rangle - \left\langle S \right\rangle^{2}\\
			\end{aligned}
\end{equation}
Thus, the baryon-strangeness correlation coefficient:
\begin{equation}
\begin{aligned}
C_{BS} = -3\frac{C(\delta B,\delta S)}{C(\delta S,\delta S)} = -3\frac{\left\langle BS \right\rangle - \left\langle B \right\rangle \left\langle S \right\rangle}{\left\langle S^{2}  \right\rangle - \left\langle S \right\rangle^{2}}
			\end{aligned}
\end{equation}
\end{appendices}

	  \begin{appendices} 
      \section{The statistical error of $C_{BS}$} 
      In the appendix of Ref.~\cite{Luo_UrQMD}, the authors showed in detail how to calculate the statistical uncertainty by way of the covariance of the multivariate moments.
According to the definition of the covariance,
	 \begin{equation}
\operatorname{cov}\left(f_{i, j}, f_{k, h}\right)=\frac{1}{N}\left(f_{i+k, j+h}-f_{i, j} f_{k, h}\right),
\end{equation}
higher-order terms must be introduced for calculating the covariance.
	 So we give all the items since those are necessary for calculating the error throughout the table below. 
	 Form the equation we know the error is proportional to $1/\sqrt{N}$, however, the corresponding event statistics we use are relatively small, 
	and the statistical errors of results would be large.
	 	\begin{table*}[]
\scriptsize
\centering
\caption{This table list all the variables needed to calculate the results and statistical errors in Fig.~\ref{Fig2_Sys_E_scan_CBS}(a) at $\sqrt{s_{NN}} = 200$ GeV with hadron rescattering process in the AMPT framework.}
\label{error_info}

	\begin{tabular}{|c|c|c|c|c|c|c|c|c|c|c|c|c|c|}
\toprule
System & Events
&$\left\langle B \right\rangle$ & $\left\langle S \right\rangle$
&$\left\langle BS \right\rangle$  & $\left\langle S^{2} \right\rangle$ 
&$\left\langle S^{3} \right\rangle$  &$\left\langle S^{4} \right\rangle$
&$\left\langle BS^{2} \right\rangle$  &$\left\langle BS^{3} \right\rangle$
&$\left\langle B^{2} \right\rangle$  &$\left\langle B^{2}S \right\rangle$
&$\left\langle B^{2}S^{2} \right\rangle$  & error
      \\
\hline
$\mathrm{\leftidx{^{10}}B} + \mathrm{\leftidx{^{10}}B}$		&69800&0.0736533&0.0896132&-0.359728&1.49643&0.513395&9.36554&0.00531519&-2.32606&1.09706&0.0775501&2.47291&0.0233956 \\
$\mathrm{\leftidx{^{12}}C} + \mathrm{\leftidx{^{12}}C}$		&99800&0.0812625&0.109599&-0.493968&1.95445&0.898196&14.9372&-0.0448697&-3.89148&1.45303&0.119419&4.09048&0.0178936\\
$\mathrm{\leftidx{^{16}}O}+\mathrm{\leftidx{^{16}}O}$		&100000&0.12687&0.16194&-0.66834&2.72044&1.69218&27.0776&-0.00802&-6.71508&2.02597&0.2679&7.38878&0.0153406 \\
$\mathrm{\leftidx{^{20}}Ne}+\mathrm{\leftidx{^{20}}Ne}$		&20000&0.21585&0.20645&-0.90485&3.62225&2.55395&45.4021&0.18435&-11.515&2.81145&0.28175&13.1649&0.0283403\\
$\mathrm{\leftidx{^{40}}Ca}+\mathrm{\leftidx{^{40}}Ca}$		&20000&0.57985&0.49615&-1.86935&8.62085&12.5274&234.946&2.55565&-50.8957&6.67815&1.27765&67.8139&0.00965114 \\
$\mathrm{\leftidx{^{96}}Zr}+\mathrm{\leftidx{^{96}}Zr}$ 		&19900&2.08171&1.28734&-3.71834&24.8124&93.0328&1867.91&34.6337&-295.664&22.621&3.92558&588.743&0.0124063\\
$\mathrm{\leftidx{^{197}}Au}+\mathrm{\leftidx{^{197}}Au}$		&10000&5.8683&2.5061&-0.6767&58.6817&434.255&10679.6&258.236&-324.446&77.1819&14.6115&4089.6&0.496962\\
\bottomrule
\end{tabular}
\end{table*}

\end{appendices} 
	\end{CJK*}	
		\bibliography{no}

\end{document}